\documentclass{iau} 
\usepackage{graphicx}

\title[JD 11.~~Black holes formed by implosion of massive stars] 
{Black holes formed by direct collapse: observational evidences}

\author[I.F. Mirabel]   
{I.F. Mirabel$^{1,2}$}

\affiliation{$^1$Institute of Astronomy and Space Physics. CONICET - Universidad de Buenos Aires,   
Ciudad Universitaria, Av. Cantilo S/N , 1428 Buenos Aires - Argentina  \\ email: {\tt mirabel@iafe.uba.ar} 
\\[\affilskip]$^2$Laboratoire AIM-Paris-Saclay, CEA/DSM/Irfu−CNRS, CEA-Saclay, pt courrier 131, 91191 Gif-sur-Yvette, France \\email: {\tt felix.mirabel@cea.fr}}

\pubyear{2017}
\volume{324}  
\jname{New Frontiers in Black Hole Astrophysics}
\editors{Andreja Gomboc, \& Carole Mundell, eds.}
\begin{document}

\maketitle

\begin{abstract}
Binary black holes as the recently detected sources of gravitational waves can be formed from massive stellar binaries in the field or by dynamical interactions in clusters of high stellar density, if the black holes are the remnants of massive stars that collapsed without natal kicks that would disrupt the binary system or eject the black holes from the cluster before binary black hole formation. Here are summarized and discussed the kinematics in three dimensions of space of five Galactic black hole X-ray binaries. For Cygnus X-1 and GRS 1915+105 it is found that the black holes of $\sim$15 M$_{\odot}$ and $\sim$10 M$_{\odot}$ in these sources were formed in situ, without energetic kicks. These observations suggest that binary black holes with components of $\sim$10 M$_{\odot}$ may have been prolifically produced in the universe. 

\end{abstract}


\firstsection 
\section{Introduction}

It is assumed that a binary black hole (BBH) with components of 30-40 M$_{\odot}$ as the source of gravitational waves GW150914 (\cite[Abbott et al. 2016a]{Abbott_etal16a}), can be formed from a relatively isolated binary of massive stars if both black holes (BHs) are formed by implosion, namely, by complete or almost complete collapse of massive stars with no energetic supernovae (SNe) accompanied by a sudden large mass loss that would significantly reduce the mass of the compact objects, and in most cases unbind the binary system (\cite[Belczynski et al. 2016]{Belczynski_etal16}). BBHs can also be formed by dynamical interactions in the high stellar density environments (\cite[Rodriguez et al. 2016]{Rodriguez_etal16}) of typical globular clusters of $\leq$10$^7$ M$_{\odot}$, if the BHs are formed with no energetic kicks that will eject the BHs out from the cluster before BBH formation. Theoretical models set progenitor masses for BH formation by implosion, but observational evidences have been elusive.  

The question on how stellar black holes are formed is of topical interest for the incipient Gravitational-Wave Astrophysics. Whether BHs are formed through energetic natal supernova kicks or by implosion will determine the final evolutionary stage of massive stars, the numbers of BBHs that would be formed, and therefore the merger rates of BBHs that will be detected by gravitational-wave (GW) observations with LIGO, VIRGO and other GW collaborations. In fact, from population synthesis models it is inferred that the disruption rate of massive stellar binaries increases by $\sim$20 varying from the assumption of BH formation with no kicks to that of a kick distribution typical of neutron stars (\cite[Dominik et al. 2012]{Dominik_etal12}). Besides, the scape velocity from a typical globular cluster of $\leq$10$^7$ M$_{\odot}$ is few tens of km s$^{-1}$, and  if formed with a kick distribution typical of neutron stars most BHs would be kicked out from typical globular clusters. 

The kinematics of BH X-ray binaries can provide clues on the formation of BHs. If a compact object is accompanied by a mass-donor star in an X-ray binary, it is possible to determine the distance, proper motion, and radial velocity of the center of mass of the system, from which can be derived the velocity in three dimensions of space, and in some cases, infer the site of birth of the BH. Here are summarized the recently improved determination, mostly by VLBI at radio wavelengths, of the kinematics of five Galactic BH X-ray binaries.   

\section{Black holes formed by implosion}

\begin{figure}[h]
\begin{center}
 \includegraphics[width=5.3in]{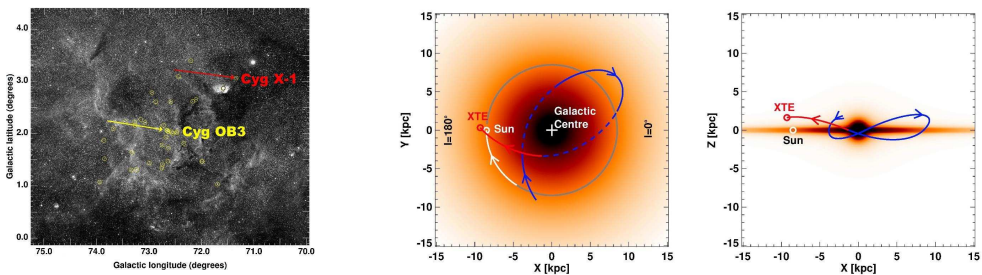} 
 \caption{\textbf{Left}: Optical image of the sky around the BH X-ray binary Cygnus X-1 and the association of massive stars Cygnus OB3. The red and yellow arrows show the magnitudes and directions of the motion in the plane of the sky of the radio counterpart of Cygnus X-1 and the average Hipparcos motion of the massive stars of Cygnus OB3 (circled in yellow) for the past 0.5 Millions years, respectively. After the formation of the BH, Cygnus X-1 remained anchored in the parent association of massive stars Cygnus OB3.   From \cite[Mirabel \& Rodrigues (2003)]{MirabelRodrigues 2003}.\hspace{30pt} \textbf{Right}: Schematic Galactic orbit of the runway BH X-ray binary XTE J1118+480 (blue curve) during the last orbital period of the Sun around the Galactic Center (240 Myr). The last section of the orbit since the source left the plane 37 $\pm$ 5  Myrs ago at a galactocentric distance of 4 $\pm$ 0.5 kpc is shown in red. The trajectory of the Sun during the later time is indicated by the thick white arc. The source left the plane towards the northern Galactic hemisphere with a galactocentric velocity of 348 $\pm$ 18 km s$^{-1}$, which after subtraction of the velocity vector due to Galactic rotation, corresponds to a peculiar space velocity of 217 $\pm$ 18 km s$^{-1}$  relative to the Galactic disk frame, and a component perpendicular to the plane of 126 $\pm$ 18 km s$^{-1}$. The galactic orbit of XTE J1118+480 has an eccentricity of 0.54. At the present epoch XTE J1118+480 is at a distance from the Sun of only 1.9 $\pm$ 0.4 kpc flying through the Galactic local neighborhood with a velocity of 145 km s$^{-1}$. Left, view from above the Galactic plane; Right, side view. From Mirabel et al. (2001)}

   \label{fig1}
\end{center}
\end{figure}

\textbf{Cygnus X-1} is an X-ray binary at a distance of 1.86 $\pm$ 0.1 kpc  composed by a BH of 14.8 $\pm$ 1.0 M$_{\odot}$  and a 09.7 lab donor star of 19.2 $\pm$1.9 M$_{\odot}$ with an orbital period of 5.6 days and eccentricity of 0.018 $\pm$ 0.003.  Cygnus X-1 appears to be at comparable distance and moving together with the association of massive stars Cygnus OB3 (\cite[Mirabel \& Rodrigues 2003)]{MirabelRodriges 2003}. Therefore, it has been proposed that the BH in Cygnus X-1 was formed in situ and did not receive an energetic trigger from a natal or nearby supernova. The recent VLBI measurements of the parallax and proper motion of Cygnus X-1  (\cite[Reid et al. 2011]{Reid_etal11}), and the new reduction of the Hipparcos data to infer the mean distance and proper motion of Cygnus OB3, show that within the errors, the distance and proper motion, -as well as the radial velocity of the black hole X-ray binary barycenter-, are consistent with those of the association of massive stars, which reaffirmed the conjecture that Cygnus OB3 is the parent association of Cygnus X-1. The motions on the plane of the sky of Cygnus X-1 and Cygnus OB3 are shown on the left side of  Figure \ref{fig1}. 

The upper limit of the velocity in three dimensions of Cygnus X-1 relative to the mean velocity of Cygnus OB3 is 9 $\pm$ 2 km s$^{-1}$, which is typical of the random velocities of stars in expanding stellar associations. From the equations for spherical mass ejection in BH formation  (\cite[Nelemans et al. 1999]{Nelemans_etal99}) it is estimated that the maximum mass that could have been suddenly ejected to accelerate the binary without disruption to a velocity of 9 $\pm$ 2 km s$^{-1}$ is less than 1 $\pm$ 0.3 M$_{\odot}$. Indeed, there are no observational evidences for a SN remnant in the radio continuum, X-rays, and atomic hydrogen surveys of the region where Cygnus X-1 was most likely formed. It was estimated that the initial mass of the progenitor is more than 40 $\pm$ 5 M$_{\odot}$ and that it may have lost $\sim$6 M$_{\odot}$ by stellar winds during a Wolf-Rayet stage (\cite[Mirabel \& Rodrigues (2003)]{MirabelRodrigues 2003}. The lower mass limit of $\sim$40 M$_{\odot}$ for the progenitor is the same as the mass theoretically predicted for the transition from fall-back to complete collapse for a BH progenitor of solar metallicity (\cite[Fryer et al. 2012]{Fryer_etal12}).

\textbf{GRS 1915+105} is a low-mass X-ray binary containing a BH of 10.1 $\pm$  0.6 M$_{\odot}$ and a donor star of spectral type K-M III of 0.5 $\pm$  0.3 M$_{\odot}$ with a 34 day circular orbital period. The companion overflows its Roche lobe and the system exhibits episodic superluminal radio jets. Using a decade of astrometry with the NRAO Very Long Baseline Array it was measured a parallax distance of 8.6 $\pm$ 1.8 kpc and a proper motion for GRS 1915+105, that together with the published radial velocity of the binary barycenter, it is inferred a modest peculiar velocity of 22 $\pm$  24 km s$^{-1}$  (\cite[Reid et al. 2014]{Reid_etal14}), which is consistent with the earlier proposition that the BH in GRS 1915+105 was formed without a strong natal kick (\cite[Mirabel et al. 2001]{Mirabel_etal01}). 
The modest peculiar speed of 22 $\pm$ 24 km s$^{-1}$  and a donor star in the giant branch suggest that GRS 1915+105 is an old system that has orbited the Galaxy many times, acquiring a peculiar velocity component on the galactic disk of 20-30 km s$^{-1}$, consistent with the velocity dispersions of $\sim$20 km s$^{-1}$ of old stellar systems in the thin disk, due to galactic diffusion by random gravitational perturbations from encounters with spiral arms and giant molecular clouds. 

The kinematics in three dimensions of Cygnus X-1 relative to the parent association of massive stars Cygnus OB3, and the kinematics of GRS 1915+105 relative to its Galactic environment, suggest, irrespective of their origin in isolated massive stellar binaries or in typical dense clusters of less than 10$^7$ M$_{\odot}$, that the black holes in the x-ray binaries Cygnus X-1 and in GRS 1915+105, were formed in situ by complete or almost complete collapse of massive stars, with no energetic kicks from natal supernovae. These observational results are consistent with theoretical models, and in particular, with one of the most recent models, where the black holes of $\sim$15 M$_{\odot}$  in Cygnus X-1, and of $\sim$10 M$_{\odot}$ in GRS 1915+105, may be formed by complete or almost complete collapse of stellar helium cores (\cite[Sukhbold et al. 2016]{Sukhbold_etal16}).

\section{Runaway black hole x-ray binaries}

\textbf{GRO J1655-40} is an X-ray binary with a BH of 5.3 $\pm$  0.7 M$_{\odot}$ and a F6-F7 IV donor star with a runaway velocity of 112 $\pm$ 18 km s$^{-1}$ moving in a highly eccentric (e=0.34 $\pm$ 0.05) Galactic orbit (\cite[Mirabel et al. 2002]{Mirabel_etal02}). The runaway linear momentum of the X-ray binary is similar to that of a solitary neutron star with a runaway velocity of $\sim$420 km s$^{-1}$. 

\textbf{XTE J1118+480} is a high-galactic-latitude (l=157$^{o}$.78, b=+62$^{o}$.38) x-ray binary with a BH of 7.6 $\pm$ 0.7 M$_{\odot}$ and a 0.18 M$_{\odot}$ donor star of spectral type K7 V to M1 V, moving in a highly eccentric orbit around the Galactic center region, as some ancient stars and globular clusters in the halo of the Galaxy. On the right two panels of Figure 1 are shown the top and side views of the orbital path of this X-ray binary relative to the Galactic disk (\cite[Mirabel et al. 2001]{Mirabel_etal01}).

\textbf{V404 Cyg (GS 2023+338)} is a low mass X-ray binary system composed of a BH of 9.0 $\pm$ 0.6 M$_{\odot}$  and a 0.75 M$_{\odot}$ donor of spectral type K0 IV. From astrometric VLBI observations, it was measured for this system a parallax that corresponds to a distance of 2.39 $\pm$ 0.14 kpc and proper motion (\cite[Miller-Jones et al. 2009a]{Miller-Jones_etal09a}), from which it is derived a peculiar velocity of 39.9 $\pm$ 5.5 km s$^{-1}$, with a component on the Galactic plane of 39.6 km s$^{-1}$ (\cite[Miller-Jones et al. 2009b]{Miller-Jones_etal09b}), that is $\sim$2 times larger than the expected velocity dispersion in the Galactic plane. 

\textbf{Discussion on the origin of the runaway kinematics of black hole X-ray binaries:}
GRO J1655-40 and V404 Cyg are in the Galactic disk (b $\leq$ 3.2$^{o}$; z $\leq$ 0.15 kpc) where likely were formed, except XTE J118+480 which is in the Galactic halo (b = 62.3$^{o}$; z = 1.5 kpc) and could either have been propelled to its present position from the Galactic disk by a supernova explosion, or have been formed in a globular cluster from which it could have escaped with a mild velocity of tens km s$^{-1}$. If these X-ray binaries were formed from binary stars in a field of relative low density, one would conclude that the three BHs with $\leq$10 M$_{\odot}$ were formed with significant natal kicks, whereas the BHs with $\geq$ 10 M$_{\odot}$ were formed with no energetic supernovae. But if the runaway BH X-ray binaries were formed in dense stellar clusters, the anomalous velocities of the X-ray binaries barycenter could have been caused by dynamical interactions in the stellar cluster, or by the explosion of a nearby massive star, rather than by the explosion of the progenitor of the runaway BH. In this context, the runaway velocities and supernova nucleosynthetic products in the atmospheres of the donor stars in GRO J1655-40, XTE 1118+480 and V404 Cyg could be due to the explosion of a nearby star, rather than to the explosion of the BH progenitor. In fact, the three BH runaway X-ray binaries have low mass donors, and their linear momenta are comparable to those of runaway massive stars ejected from multiple stellar systems by the Blaauw mechanism. Therefore, without knowing the origin of a runaway X-ray binary, it is not possible to constrain from its peculiar velocity alone, the strength of a putative natal supernova kick to the compact object in the x-ray binary. It is expected that future parallax distances and proper motions of BHs and their environment, determined at radio wavelengths by VLBI and at optical wavelengths with GAIA, will provide further observational constraints on the physical mechanisms of stellar BH formation.   

\section{Conclusion}

• Stars of solar metallicity and $\sim$40 M$_{\odot}$ may collapse directly to form black holes without energetic supernova explosions. The kinematics of Cygnus X-1 shows that the black hole of $\sim$15 solar masses did not receive a strong natal kick, and was formed by implosion from a progenitor of $\sim$40 M$_{\odot}$, that  by exchange of mass with the binary companion and stellar winds in a Wolf-Rayet phase lost $\sim$25 M$_{\odot}$. The kinematics of GRS 1915+105 shows that its black hole of $\sim$10 M$_{\odot}$ was also formed in situ with no strong natal kick. Recent core-collapse models of massive stars based on neutrino-powered explosions are consistent with these observations.

• The formation by implosion of black holes of $\sim$10 M$_{\odot}$ in X-ray binaries is consistent with the observation of GW151226, produced by fusion of black holes of $\sim$10 M$_{\odot}$ (\cite[Abbott et al. 2016b]{Abbott_etal16b}). 

• The interpretation of the runaway velocity of an X-ray binary as a kick velocity from a natal supernova imparted to the compact object in the runaway binary is uncertain without knowledge of the binary origin. X-ray binaries may be originated in different environments and their runaway velocities be caused by a variety of physical mechanisms.

• Parallax distances and proper motions of black hole X-ray binaries determined by VLBI observations at radio wavelengths, and with GAIA at optical wavelengths, will allow to determine accurate velocities in three dimensions for larger samples of sources, and therefore, will allow to better constraint models on stellar black hole formation. 

• A short review on the current optical and infrared observations that may provide insights on the nature of the stellar progenitors of black holes can be found in Mirabel (2017).

\end{document}